# Plans for the Integration of *grid tools* in the CMS computing environment

C. Grandi
*INFN Bologna, Viale Berti-Pichat 6/2, 40127 Bologna, ITALY*
*on behalf of the CMS Computing and Core Software group*

The CMS collaboration has a long term need to perform large-scale simulation efforts, in which physics events are generated and their manifestations in the CMS detector are simulated. Simulated data are then reconstructed and analyzed by the physicists to support detector design and the design of the real-time event filtering algorithms that will be used when CMS is running. Up to year 2002 the distribution of tasks among the different regional centers has been done mainly through manual operations, even though some tools for data transfer and centralization of the book-keeping were developed. In 2002 the first prototypes of CMS distributed productions based on grid middleware have been deployed, demonstrating that it is possible to use them for real data production tasks. In this work we present the plans of the CMS experiment for building a production and analysis environment based on the grid technologies in time for the next big Data Challenge, which is foreseen for the beginning of year 2004.

## 1. CMS DATA CHALLENGES ON THE GRID IN 2002

During 2002 the integration of the CMS production environment with grid tools has been carried on in the USA and in Europe addressing complementary issues. This is mainly due to the fact that it is still missing the possibility to submit DAG's (Direct Acyclic Graphs) through the European DataGrid Resource Broker.

The USCMS Integration Grid Test-bed (IGT) followed a *bottom-up* approach. It was based on the Virtual Data Toolkit (VDT), using DAGMan and Condor-G as front-end. MOP was used to produce DAG's that were submitted to the test-bed. About 1.5 million events of the official CMS production were produced on the IGT in about one month. The whole production has been managed by less than 2 FTE's.

In Europe a stress test was performed on the European DataGrid (EDG) test-bed using the high-level tools provided by EDG. The CMS production tools IMPALA and BOSS were interfaced to the EDG middleware, exploiting the Resource Broker for resource location and the Replica Manager for data location and management. More than 250,000 events of the official CMS productions were produced during about 3 weeks. This *top-down* approach showed that the CMS computing system might benefit from the use of these high-level tools, but the stability has to be improved in order to increase the efficiency and reduce the number of FTE's needed to manage the production.

More details about the CMS data challenges using grid tools may be found in [1].

## 2. COMING CMS DATA CHALLENGES

The next important milestone for CMS computing is the Data Challenge in 2004 (DC04), also known as *5% Data Challenge*. The 5% refers to a fraction of a final, 100% full-luminosity computing configuration. That corresponds to about 25% of the complexity required for initial LHC running in 2007. The emphasis of the challenge is on the validation of the deployed grid model on a sufficient number of Tier-0, Tier-1, and Tier-2 sites. With DC04 CMS intends to perform a large-scale test of the computing and analysis models themselves. Thus a six-months pre-challenge period is anticipated in the second half of 2003 (Pre-Challenge Production, PCP03), comprising the simulation and the digitization of the data samples at the different CMS Regional Centers.

The challenge itself consists of the reconstruction and selection of the data at the T0 (Tier-0 computing center at CERN), with distribution to the T1/T2 sites and synchronous analysis. It should also be based on GEANT4 as the event simulation toolkit and on the new LCG Persistency framework, based on POOL and ROOT. Details about the CMS analysis framework may be found in [2].

DC04 is a "pure" computing challenge. For this reason CMS is committed to use the grid-enabled environment that will be set up by LCG (LHC Computing Grid) Project (LCG-1 test-bed) for the Data Challenge itself. On the other hand distribution of tasks to the Regional Centers during the pre-challenge production will be done manually. An increasing fraction of the pre-challenge production is expected to be done on the LCG-1 as soon as the stability increases.

The subsequent challenges will take place in 2005 and 2006, and are scaled in turn to be 50% and 100% of LHC turn-on complexity.

## 3. THE 2003 PRE-CHALLENGE PRODUCTION (PCP03)

### 3.1. Resource estimation

To reach the requested scale of 5% for the 2004 Data Challenge, a sample of 50 million events will be produced. The summary of resources needed at the various CMS Regional Centers for the different steps are reported in table 1 assuming 550 Si2000 CPU's. The resources needed for the reconstruction at the Tier-0 during DC04 are also reported, assuming 700 Si2000 CPU's.

|  | **Simulation** | **Digitization** | **Reconstruction** |
|---|---|---|---|
| CPU per event | 160 KSi2K·s | 8 KSi2K·s | 12 KSi2K·s |
| Total CPU | 3086 KSi2K·month | 150 KSi2K·month | 230 KSi2K·month |
| Output per event | 2 MB | 1.5 MB | 0.5 MB - ESD 20 KB - AOD |
| Total size of sample | 100 TB | 75 TB | 26 TB |
| Resource request | ~1000 CPU for 5 months | ~150 CPU for 2 months | ~460 CPU for 1 months |

Table 1: Resources needed for PCP03 and for the DC04 reconstruction.

The pre-challenge phase includes also the transfer of the digitized samples from the production sites to CERN, which implies a 1 TB/day transfer rate sustained over 2 months.

### 3.2. Boundary conditions for PCP03

The pre-challenge production will start in a period that is particularly critical for the CMS computing system.

The CMS persistency system is changing: POOL, produced by the LCG project, is replacing Objectivity/DB. This implies also a change in the C++ compiler: gcc 3.2 is replacing gcc 2.95.2. This issue is further complicated by the fact that the grid middleware is being delivered by the LCG using the gcc 2.95.2 compiler. For this reason it will not be possible to use C++ API from within CMS applications.

The operating system is changing from Red Hat 6.1 to Red Hat 7.3. The change is needed in order to be able to exploit the new hardware available on the market, which doesn't support Red Hat 6.1.

The grid middleware will be delivered by the LCG project using a new structure, which consists of EDG middleware released on top of VDT.

### 3.3. Strategy for PCP03

The pre-challenge production cannot fail because the data are needed to start the data challenge itself. For this reason the basic strategy for the CMS production team is to run on dedicated, fully controllable resources without the need of grid tools. Nevertheless CMS needs to gain experience in the use of grid tools for the DC04, so grid-based prototypes will be developed, but they have to be compatible with the basic non-grid environment. To accomplish with these boundary conditions the strategy is:
- The CMS production tools must be modular, so that it will be possible to produce jobs that can run in different environment;
- The produced jobs should make as little assumptions on the runtime environment as possible, i.e. run like in a *sandbox*;
- The system should allow monitoring of the status of the jobs, possibly also while the jobs are running.

### 3.4. CMS production tools

McRunJob [3] is a tool for job preparation that is modular. It already has plug-in's for reading job creation instructions from:
- *RefDB*, which is the CMS reference database, where CMS production requests are placed (see below);
- A simple graphical interface.

It has a plug-in for submitting jobs to:
- A local resource manager.

Others are in preparation that can submit to:
- DAGMan/Condor-G (like in the grid-production that has been done in 2002 by USCMS);
- The EDG Resource Broker (like in the grid-production that has been done in 2002 on the EDG test bed);
- The Chimera system [4], i.e. producing *transformations* in the Virtual Data Language.

This is shown in figure 1.

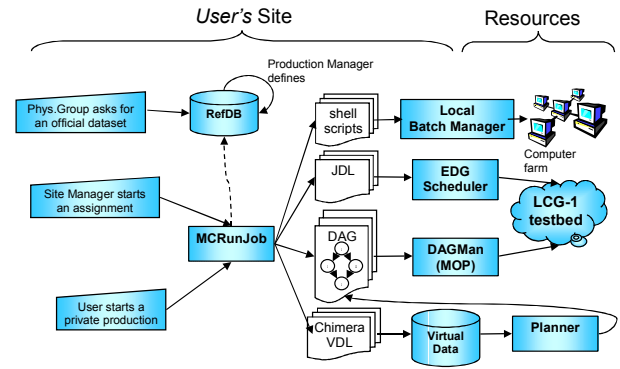

Figure 1: Hybrid Production model for Pre-Challenge Production in 2003: MCRunJob is a modular tool that allows preparing jobs for different running environments.

Jobs will be prepared in such a way that when they start they'll have local input data and XML POOL catalogue. It will be responsibility of the CMS production software to execute the needed file transfer operations in advance. Furthermore the job will write output data, as well as dataset and job metadata, locally; they will be moved to their final destinations asynchronously (at the end of the job or on an explicit request by the production manager). To increase the real-time monitoring of the production operations, synchronous components will optionally update central catalogues. If they fail the job will continue and the catalogues will be updated asynchronously. Figure 2 describes this *limited sandbox* environment.

All file transfers between the user site or a remote storage and the worker node (i.e. the node where the job is actually executed) are controlled by the CMS production software, which will optionally use external tools (e.g. EDG Job Submission System, additional DAG nodes, etc...).

The program that is submitted to the local or grid scheduler is not simply the user job, i.e. the CMS application, but rather a wrapper. The job wrapper starts not only the user job but also one or more processes that read and interpret the job output and try to update the remote catalogues. A first optional process produces a list of updates that are stored in a journal file. A second process is the one that tries to do the remote updates. An important concept is that the user job is completely decoupled from the *remote updater* so that if it fails, the job can continue (and finish) without delays. If it finishes without being able to do all the updates, it is always possible to do them asynchronously from the journal file that is transferred back to the submitter together with the job output.

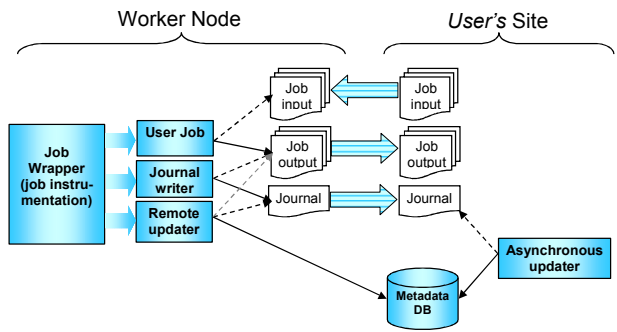

Figure 2: *Limited sandbox* environment for jobs running during the Pre-Challenge Production 2003. Continuous and dashed arrows represent push and pull of information respectively. Detailed description is in the text.

The kinds of information that CMS needs to store in the central databases are basically two: job metadata and dataset metadata.

Job metadata are parameters that represent the job running. The job metadata database should be able to answer questions like "*when did the job start?*", "*is it finished?*", but also "*how many events did it produce so far?*", i.e. it should be able to handle also application-specific metadata. BOSS [5] is a CMS-developed system that does this extracting the information from the job standard input/output/error streams through a set of filter processes provided by the user. The default remote updater is based on native MySQL calls, but the user may build others. A remote updater based on R-GMA is being developed to make the system robust in a distributed environment. Scalability tests are being done at the time of writing.

Dataset metadata are parameters that provide both the instructions needed to produce the dataset and the details about the production process. The system used by CMS is the *RefDB* [6]. The system has a web interface used by the CMS physics groups to place production requests, a central database (at CERN) where the requests are stored, and a set of tools needed by the CMS site manager to retrieve instructions and store results. The system is able to answer questions like "*by what (logical) files is a dataset made of?*", but also "*what input parameters to the simulation program where (or have to be) used?*", "*how many events of this dataset have been produced so far?*". Parameters may be updated in the RefDB in many ways: by a manual Site Manager operation, by an automatic e-mail sent by the job at the end of running. A remote updater similar to BOSS + R-GMA is being developed for running in a grid environment. Mapping of logical names to GUID (Grid Unique IDentifiers) and of GUID to physical file names will be done on the grid by the Replica Manager.

Production software is in general delivered to the CMS Regional Centers as `rpm` or as DAR (Distribution After Release, a system based on unix `tar` to install and configure CMS software). Grid-enabled sites will either get the software as `rpm` (as part of the CMS private software) or using PACMAN (a *package manager* first developed by the ATLAS experiment). In both cases software is pre-installed. The possibility to install the software *on-demand* at the time the job starts is under investigation. In any case the installation has to be advertised so that information systems know about which sites have the software installed and what don't.

Event digitization has special requirements from the point of view of data access. A special data sample (*pile-up* data) is used to superimpose a number of background events to the event to be digitized, thus stressing the server where it is stored. A careful partitioning of computing farms is needed so that the correct number of processes accesses the same pile-up servers at the same time. This can be achieved either by running the digitization step on well controlled, non-grid resources, or by distributing the pile-up sample as part of the CMS production software. But this second solution has to deal with the size of the pile-up sample, which is of about 100 GB.

CMS plans to do most data transfers using grid tools: the Replica Manager or directly gridFTP and SRB (Storage Resource Broker). Some sites doing non-grid productions may require the use of other tools; `bbftp` and `scp` have already been used in 2002 and it is foreseen to keep the possibility to use them through a CMS interface.

CMS is already testing a prototype of SRM (Storage Resource Manager) to access data on mass storage systems.

## 4. THE 2004 DATA CHALENGE (DC04)

### 4.1. DC04 Workflow

The DC04 itself will basically consist in the following activities:

- Reconstruction of digitized events at the Tier-0 at a rate corresponding to the 5% of the rate of LHC running at full luminosity (25 Hz, 50 MB/s); The reconstruction process produces Event summary Data (ESD) and Analysis Object Data (AOD).

- Distribution of AOD data to all the Tier-1 centers and distribution of ESD data to at least one Tier-1 center. Also digitized (raw) data are supposed to be distributed to at least one tier-1 center, but since the CERN bandwidth cannot cope with this data transfer, a copy of the raw data is stored in all Tier-1 before DC04 starts.
- Definition of *Express lines* and *calibration streams* to be transferred to selected Tier-1 centers.
- Archiving of raw data to CERN tape library.
- Analysis of the express lines at the selected Tier-1 centers.
- Re-calculation of the calibration constants from the calibration streams in at least one Tier-1 center; distribution of the updated *conditions database* to the Tier-0 and to the other Tier-1 centers.
- Re-processing of digitized events using the updated conditions database;
- Analysis of AOD, ESD and occasionally of raw data at the Tier-2 and Tier-1 centers.

### 4.2. DC04 Strategy

DC04 is a computing challenge: CMS is committed to use the LCG-1 resources and services (possibly integrated by other CMS resources).

It is foreseen to use:

- the Replica Manager services to locate and move the data;
- the Workload Management System to find resources for running the jobs and storing the data;
- a Grid-wide monitoring system;
- client-server tools for analysis, e.g. Clarens [7].

Actually the strategy that CMS will follow for DC04 will be determined by the results it will get from the grid-prototypes during the pre-challenge production.

### 5. SUMMARY

The next CMS computing challenges will be done in a very dynamic environment. In particular the Data Challenge in 2004 will be done on the LCG-1 test-bed, which is not completely determined at the time of writing.

The Pre-Challenge Production, which will be done in the second half of 2003, is already well defined. Since it cannot fail but at the same time CMS needs to gain experience in the use of grid tools in view of the data challenge itself, it is important to use flexible production tools that may run both in a local or in a distributed environment. The pre-production will be done basically outside the Grid but will provide an ideal proof of maturity for Grid-based prototypes.

The Data Challenge architecture will be built on the experience CMS will gain during the pre-production.

### References

[1] P. Capiluppi, "*Running CMS software on Grid Testbeds*", CHEP03, San Diego, March 2003.
[2] V. Innocente, "*CMS Data Analysis: Current Status and Future Strategy*", CHEP03, San Diego, March 2003.
[3] G. Graham, "*McRunJob: A Workflow Planner for Grid Production Processing*", CHEP03, San Diego, March 2003.
[4] R. Cavanaugh, "*Virtual Data in CMS Production*", CHEP03, San Diego, March 2003.
[5] C. Grandi, "*BOSS: a tool for batch job monitoring and book-keeping*", CHEP03, San Diego, March 2003.
[6] V. Lefebure, "*RefDB: aReference Database for CMS Monte Carlo Production*", CHEP03, San Diego, March 2003.
[7] C. Steenberg, "*The Clarens web services architecture*", and "*Clarens client and server applications*", CHEP03, San Diego, March 2003.